\documentstyle[prl,twocolumn,aps,epsfig]{revtex}
\begin{document}

\title{Topological Chern indices in molecular spectra}

\author{F. Faure\protect\( ^{1}\protect \) and B. Zhilinskii\protect\( ^{2}\protect \)}

\address{\protect\( ^{1}\protect \) L.P.M.M.C. (Maison des Magistères Jean Perrin,
CNRS) BP 166, 38042 Grenoble Cedex 9, France. \\
 faure@labs.polycnrs-gre.fr \\
 \protect\( ^{2}\protect \) Universit\'{e} du Littoral, MREID, 145 av. M.
Schumann, 59140 Dunkerque , France \\
 zhilin@univ-littoral.fr }

\date{\today{}}

\maketitle
\draft

\begin{abstract}
Topological Chern indices are related to the number of rotational states in
each molecular vibrational band. Modification of the indices is associated to
the appearance of ``band degeneracies'', and exchange of rotational states
between two consecutive bands. The topological dynamical origin of these indices
is proven through a semi-classical approach, and their values are computed in
two examples. The relation with the integer quantum Hall effect is briefly discussed. 
\end{abstract}

\newcommand{\R}{I\! \! R}

\newcommand{\C}{I\! \! \! \! C}

\newcommand{\Z}{Z\! \! \! Z}

\newcommand{\N}{I\! \! N}

\bigskip

Topological numbers play an important role in many area of physics \cite{thouless2},
but their appearance in molecular physics and especially in rovibrational problems
has not been systematically appreciated so far. Simple molecular systems typically
allow adiabatic separation of vibrational and rotational motion. For non-degenerate
isolated electronic state (this is the case of ground state for most molecules)
the rovibrational energy level system consists of vibrational bands, each associated
with one or several degenerate vibrational states. If the rovibrational coupling
is not too strong, further splitting of the rovibrational structure into sub-bands
can be clearly seen. The well-known example is the splitting of the triply degenerate
vibrational structure for a spherically symmetrical molecule into three sub-bands
due to the first-order Coriolis interaction \cite{Landau,BiL}. Within each
sub-band formed by \( 2j+1-C \) levels, with respectively \( C=+2,0,-2 \),
all energy levels are usually characterized by the quantum number \( j \) of
the total angular momentum, and by another quantum number \( R=j+C/2 \) which
characterizes the coupling of \( j \) with the vibrational angular momentum.

In this letter we show that the integer \( C \) can be defined in much more
general situation as an additional quantum number having a precise topological
meaning, namely a Chern index, whose construction will be explained below. This
index is defined in the classical limit of the rotational motion. It can be
associated with any vibrational band presented in the energy level pattern of
molecular systems. Theorem (\ref{e:formule2}) relates this topological index
to the number of rotational states within the band. A modification of the index
is associated with the formation of a contact (a degeneracy) between two consecutive
vibrational bands, and is shown to generically imply an exchange of one rotational
state between the two bands.

Such a relation was first conjectured in 1988 \cite{EurophL} after the study
of the simple model (\ref{HpolJSsimpl}), and a number of effective Hamiltonians
reconstructed from experimental data (see Ref.\cite{MolEx,ZhBr,BrZh} for the
molecular examples: SiH\( _{4} \), CD\( _{4} \), SnH\( _{4} \), CF\( _{4} \),
Mo(CO)\( _{6} \)). The universal character of the redistribution phenomena
and its relevance to integer Hall effect was discussed on several occasions
\cite{Zhil,Bel,Zeld}.

To explain the physical phenomenon and to prepare the formulation of a rigorous
statement let us consider a toy problem which involves two quantum angular momenta
\( \mathbf{J} \) and \( \mathbf{S} \), with fixed modulus \( {\mathbf{J}}^{2}=j(j+1) \)
and \( {\mathbf{S}}^{2}=s(s+1) \) with \( j,s \) integer or half-integer:
\( \mathbf{J} \) acts in the space \( {\mathcal{H}}_{j} \) of dimension \( (2j+1) \)
which is the irreducible representation space of the \( SU(2) \) group. Similarly,
\( \mathbf{S} \) acts in the space \( {\mathcal{H}}_{s} \) of dimension \( (2s+1) \).
The total space is \( {\mathcal{H}}_{\textrm{tot}}={\mathcal{H}}_{j}\otimes {\mathcal{H}}_{s} \)
with dimension \( (2j+1)(2s+1) \).

The most general quantum Hamiltonian \( \widehat{H}({\mathbf{S}},{\mathbf{J}}/j) \)
we will consider is a hermitian operator acting in \( \mathcal{H}_{\textrm{tot}} \)
and its action in space \( {\mathcal{H}}_{j} \) is supposed to be expressed
in terms of the operators \( {\mathbf{J}}/j \). The factor \( 1/j \) is introduced
here to ensure the existence of the classical limit for \( j\rightarrow \infty  \).
An extremely simple form of \( \hat{H} \) is 
\begin{eqnarray}
\hat{H}=(1-t)S_{z}\, \, +\frac{t}{j}(\mathbf{J}\cdot \mathbf{S}), & \label{HpolJSsimpl} 
\end{eqnarray}
 with \( j>s \), \( t\in \R  \), which was used initially in Ref.\cite{EurophL}
to study the redistribution phenomenon and further in Ref.\cite{PhysL} to establish
its relation with the classical monodromy. We use this Hamiltonian (\ref{HpolJSsimpl})
to illustrate the strict formulation of our result (\ref{e:formule2}), but
its validity extends to a general \( \widehat{H}({\mathbf{S}},{\mathbf{J}}/j) \).

In the two extremes limits \( t=0 \) (no ``spin-orbit'' coupling), and \( t=1 \)
(``spin-orbit'' coupling), the energy level spectrum of (\ref{HpolJSsimpl})
shows different patterns of energy levels into bands indexed by \( g \). For
\( t=0 \) all energies \( E_{g}=g \), \( g\in \{-s,..,+s\} \) appear with
the same multiplicities \( N_{g}=(2j+1) \). For \( t=1 \) the spectrum is
split into degenerate multiplets characterized by different eigenvalues of the
coupled angular momentum \( {\mathbf{N}}^{2}=({\mathbf{J}}+{\mathbf{S}})^{2} \).
As in the case of standard spin-orbit coupling with \( j>s \) there are \( 2s+1 \)
different levels \( E_{g}=[n(n+1)-j(j+1)-s(s+1)]/(2j) \), \( g=n-j\in \{-s,\ldots ,+s\} \),
with different multiplicities \( N_{g}=(2j+1)+2g. \) The two different limiting
cases for the structures of the \( (2s+1) \) bands suggest to introduce a new
quantum number \( C_{g} \) associated with the value of \( N_{g} \) within
each band. 

To define \( C_{g} \) for a general Hamiltonian \( \widehat{H}({\mathbf{S}},{\mathbf{J}}/j) \)
we assume that \( j\gg s \), so that it is physically reasonable to consider
\( {\mathbf{J}}_{cl}=\mathbf{J} \) as classical, whereas \( \mathbf{S} \)
remains quantum. The classical dynamics for \( {\mathbf{J}}_{cl} \) can be
defined through the \( SU(2) \) coherent states \( |{\mathbf{J}}_{cl}\rangle  \)
\cite{ec1}. The classical phase space for \( {\mathbf{J}}_{cl} \) is the sphere
\( S_{j}^{2} \). From \( d{\mathbf{J}}_{cl}/dt=\partial _{{\mathbf{J}}_{cl}}H_{cl}\times {\mathbf{J}}_{cl} \)
and because of the factor \( 1/j \) in Eq.(\ref{HpolJSsimpl}), \( {\mathbf{J}}_{cl} \)
corresponds to a slow dynamical variable compared to \( \mathbf{S} \), and
the Born-Oppenheimer approximation suggests to consider for each \( {\mathbf{J}}_{cl} \),
the Hermitian operator \( \hat{H}_{s}({\mathbf{J}}_{cl})=\langle {\mathbf{J}}_{cl}|\hat{H}|{\mathbf{J}}_{cl}\rangle  \)
acting on \( {\mathcal{H}}_{s} \), with spectrum \( \hat{H}_{s}({\mathbf{J}}_{cl})|\psi _{g,{\mathbf{J}}_{cl}}\rangle =E_{g,{\mathbf{J}}_{cl}}|\psi _{g,{\mathbf{J}}_{cl}}\rangle  \),
\( g\in \{-s,..,+s\} \). We suppose that for each \( {\mathbf{J}}_{cl} \),
the \( (2s+1) \) eigenvalues are isolated: \( E_{-s,{\mathbf{J}}_{cl}}<E_{-s+1,{\mathbf{J}}_{cl}}<\ldots <E_{+s,{\mathbf{J}}_{cl}} \).
This is the generic situation, because degeneracies are of codimension 3, and
\( {\mathbf{J}}_{cl}\in S_{j}^{2} \) is only two-dimensional.

For each level \( g \), let us note \( [|\psi _{g,{\mathbf{J}}_{cl}}\rangle ] \)
the eigenvector defined up to a multiplication by a phase \( e^{i\alpha } \).
The application \( {\mathbf{J}}_{cl}\rightarrow [|\psi _{g,{\mathbf{J}}_{cl}}\rangle ] \)
defines then a \( U(1) \) fiber bundle over the sphere \( S_{j}^{2} \), which
is the set of the all possible phases \( \alpha  \) for every values of \( {\mathbf{J}}_{cl} \).
The topology of this bundle is characterized by a Chern number \( C_{g}\in \Z  \)
\cite{nakahara}. \( C_{g} \) reveals the possible global twist of the fiber
of phases \( \alpha  \) over the sphere \( S_{j}^{2} \), in the same way the
well known M\"{o}bius strip is the real line fiber bundle over the circle \( S^{1} \)
with a global twist \( +1 \).
\begin{figure}
\par\centering \resizebox*{0.8\columnwidth}{!}{\includegraphics{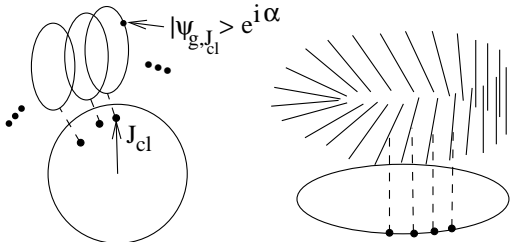} } \par{}

\caption{The Chern index \protect\protect\( C_{g}\protect \protect \) expresses the
twist made by the circles for the phases \protect\protect\( \alpha \protect \protect \)
of the eigen-vectors \protect\protect\( e^{i\alpha }|\psi _{g,{\mathbf{J}}_{cl}}\rangle \protect \protect \)
over the sphere of spin \protect\protect\( {\mathbf{J}}_{cl}\protect \protect \),
in the same way the M\"{o}bius strip has a twist made by the lines over a circle.}
\end{figure}

Note that 
\begin{equation}
\label{eq/sum}
\sum _{g=-s}^{+s}C_{g}=0,
\end{equation}
 just because of the additivity of Chern indices, and because the eigenvectors
\( (|\psi _{g,{\mathbf{J}}_{cl}}\rangle ) \) span the space \( {\mathcal{H}}_{s} \)
which does not depend on \( {\mathbf{J}}_{cl} \).

The Born Oppenheimer approach tells that the total spectrum \( \hat{H}|\phi _{i}\rangle =E_{i}|\phi _{i}\rangle ,i\in \{1,..,(2s+1)(2j+1)\} \)
in \( \mathcal{H}_{\textrm{tot}} \), can be represented as formed in \( 2s+1 \)
groups of levels (bands) with eigen-functions of each group spanning vector
spaces: \( L_{g}\subset {\mathcal{H}}_{\textrm{tot}},\quad g\in \{-s,\ldots ,+s\}. \)
We precise now the number of levels in each group:

\textbf{Theorem.}

\emph{For a general Hamiltonian \( \hat{H}({\mathbf{S}},{\mathbf{J}}/j) \),
and for \( j \) large enough then \( \dim L_{g}=\dim {\mathcal{H}}_{j}-C_{g}, \)
so}

\begin{equation}
\label{e:formule2}
N_{g}=(2j+1)-C_{g}.
\end{equation}
 The number \( N_{g} \) of states \( |\phi _{i}\rangle  \) in each band \( L_{g} \)
of \( \mathcal{H}_{\textrm{tot}} \) is given by the quantum number \( j \)
and an additional quantum number \( C_{g} \), namely the topological Chern
index.

A few remarks are in order here. Since \( (2j+1) \) is the number of quanta
in the classical phase space \( S_{j}^{2} \) for spin \( {\mathbf{J}}_{cl} \),
Eq.(\ref{e:formule2}) looks like a Weyl formula with a correction. The index
\( C_{g} \) has been defined and can be computed in ``a semi-quantal'' approach
where \( {\mathbf{J}}_{cl} \) is considered as a classical variable and \( \mathbf{S} \)
quantum. Nevertheless Eq.(\ref{e:formule2}) provides an information on the
full quantum problem: the spectrum of \( \hat{H} \). Finally the topological
nature of \( C_{g} \) reveals a qualitative and robust property of the spectrum
of \( \hat{H} \), stable under perturbations, provided no degeneracy appears
between consecutive bands. One can say that \( C_{g} \) expresses a topological
coupling between the dynamical variables \( \mathbf{J} \) and \( \mathbf{S} \).

Two approaches to the computation of the Chern indices of different bands will
be suggested. The first one we use below is algebraic. The second one uses the
Berry's connection, and is based on the curvature formula \cite{nakahara}.
This last formula could be useful for numerical computations.

The algebraic calculation is based on the geometric interpretation of the Chern
index \( C_{g} \) as the total intersection number between the one-dimensional
curve \( [|\psi _{g,{\mathbf{J}}_{cl}}\rangle ]_{{\mathbf{J}}_{cl}} \) in \( P({\mathcal{H}}_{s}) \)
with the hyperplane \( (|\psi _{0}\rangle )^{\bot }=\left\{ |\varphi \rangle \, \textrm{such}\, \, \textrm{that}\, \langle \varphi |\psi _{0}\rangle =0\right\}  \).
Here \( |\psi _{0}\rangle \in {\mathcal{H}}_{s} \) is arbitrary. Each intersection
has number \( \sigma =+1 \) (\( -1 \)) if the curve orientation is compatible
(incompatible) with the orientation of the hyperplane. \( C_{g} \) is the sum
of these intersection numbers \cite{harris1}.

\begin{figure}
\par\centering \resizebox*{0.8\columnwidth}{!}{\includegraphics{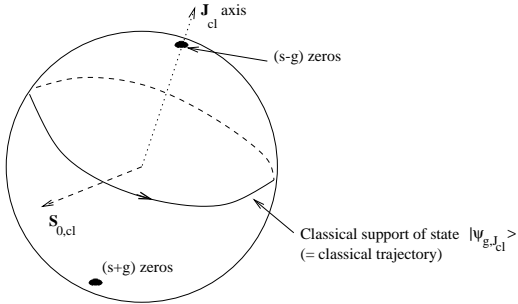} } \par{}

\caption{\label{Fig1} The Husimi distribution for the state \protect\protect\protect\( |\psi _{g,{\mathbf{J}}_{cl}}\rangle \protect \protect \protect \)
on the sphere \protect\protect\protect\( S_{s}^{2}\protect \protect \protect \)
for spin \protect\protect\protect\( {\mathbf{S}}\protect \protect \protect \).
\protect\protect\( {\mathbf{S}}_{0,cl}\protect \protect \) is a reference point. }
\end{figure}

The application of this algebraic method of calculation to the two limiting
cases of Hamiltonian (\ref{HpolJSsimpl}) is immediate. For \( t=0 \) the Hamiltonian
\( \hat{H}_{s}({\mathbf{J}}_{cl})=S_{z} \) does not depend on \( {\mathbf{J}}_{cl} \),
the application \( {\mathbf{J}}_{cl}\rightarrow [|\psi _{g,{\mathbf{J}}_{cl}}\rangle ] \)
is constant, the topology of the bundle is trivial with zero Chern index: \( C_{g}=0 \),
in accordance with Eq.(\ref{e:formule2}). For \( t=1 \) we have 
\[
\hat{H}_{s}({\mathbf{J}}_{cl})=\langle {\mathbf{J}}_{cl}|\hat{H}|{\mathbf{J}}_{cl}\rangle =\frac{1}{j}({\mathbf{J}}_{cl}\cdot {\mathbf{S}}),\]
 with \( {\mathbf{J}}_{cl} \) a vector on the sphere \( S_{j}^{2} \), and
\( {\mathbf{S}} \) a vectorial operator in \( {\mathcal{H}}_{s} \). The eigenvector
\( |\psi _{g,{\mathbf{J}}_{cl}}\rangle  \) is easily obtained from the eigenvector
\( |m_{s}=g\rangle \propto S_{+}^{(s+g)}|0\rangle  \) of \( S_{z} \) by a
rotation which transforms the \( z \) axis to the \( {\mathbf{J}}_{cl} \)
axis on the sphere. It is convenient to choose a coherent state \( |{\mathbf{S}}_{0,cl}\rangle  \)
as the reference state \( |\psi _{0}\rangle  \), where \( {\mathbf{S}}_{0,cl} \)
is an arbitrary classical spin \( \mathbf{S} \). An intersection of the curve
\( [|\psi _{g,{\mathbf{J}}_{cl}}\rangle ]_{{\mathbf{J}}_{cl}} \)with the hyperplane
\( (|\psi _{0}\rangle )^{\bot } \) is then given by the equation \( \langle {\mathbf{S}}_{0,cl}|\psi _{g,{\mathbf{J}}_{cl}}\rangle =0 \)
which has a very simple interpretation: the point \( {\mathbf{S}}_{0,cl} \)
is a zero of the Husimi representation \( \textrm{Hus}({\mathbf{S}}_{cl})=\left| \langle {\mathbf{S}}_{cl}|\psi _{g,{\mathbf{J}}_{cl}}\rangle \right| ^{2} \)
of the state \( |\psi _{g,{\mathbf{J}}_{cl}}\rangle  \) on the sphere \( S^{2}_{s} \)
(the classical phase space of \( {\mathbf{S}} \)), shown in figure \ref{Fig1}.
The Husimi representation has \( (s-g) \) zeros inside the (oriented) trajectory,
and \( (s+g) \) zeros outside \cite{zero2}. When the axis \( {\mathbf{J}}_{cl} \)
moves with direct orientation on the whole sphere, the \( (s-g) \) zeros pass
with positive orientation over the fixed point \( {\mathbf{S}}_{0,cl} \), giving
\( \sigma _{1}=s-g \), whereas the \( s+g \) zeros pass with negative orientation
over \( {\mathbf{S}}_{0,cl} \), giving \( \sigma _{2}=-(s+g) \). So \( C_{g}=\sigma _{1}+\sigma _{2}=-2g \)
in accordance with Eq.(\ref{e:formule2}).

\begin{figure}
\par\centering \resizebox*{0.8\columnwidth}{!}{\includegraphics{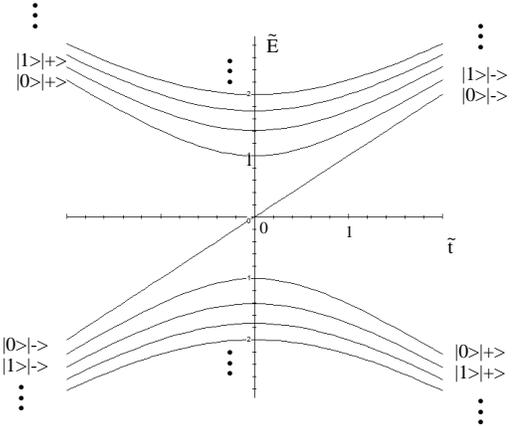} } \par{}

\caption{\label{fig:spectre}Spectrum of rotational states for the Hamiltonian (\ref{e:modele}).
\protect\( \tilde{t}=0\protect \) corresponds a conical contact (degeneracy)
between the two bands, with a topological charge \protect\protect\( \Delta C_{+}=-1\protect \protect \),
and an exchange of one rotational state, \protect\protect\( \Delta N_{+}=+1\protect \protect \).}
\end{figure}

In this last example, the variation of Chern indices \( \Delta C_{g}=-2g \)
occurs at \( t=1/2 \), with a degeneracy between the bands at \( {\mathbf{J}}_{cl}=(0,0,-j) \),
giving \( \hat{H}_{s}({\mathbf{J}}_{cl})=0 \). In the case of two bands (\( s=1/2 \),
\( g=\pm 1/2 \)) then \( \Delta C_{g=\pm }=\mp 1 \), and the two bands have
a conical contact at the degeneracy. In the vicinity of the degeneracy, and
for \( j\rightarrow \infty  \), we observe that \( [J_{-},J_{+}]=-2J_{z}\simeq 2j \),
so \( a=J_{-}/\sqrt{2j} \), \( a^{+}=J_{+}/\sqrt{2j} \) fulfilled the harmonic
oscillator commutation relations. In the basis \( |\pm \rangle =|m_{s}=\pm 1/2\rangle  \)
of \( {\mathcal{H}}_{s} \), the expression of \( \hat{H} \) can be simplified
and gives 
\begin{equation}
\label{e:modele}
\hat{H}=\frac{1}{\sqrt{2j}}\left( \begin{array}{cc}
-\tilde{t} & a\\
a^{+} & \tilde{t}
\end{array}\right) ,
\end{equation}
 with \( \tilde{t}=(2t-1)\sqrt{2j} \). We also scales the energy with \( \tilde{E}=E\sqrt{2j} \),
and note \( |n\rangle \propto a^{+n}|0\rangle  \). The stationary equation
\( \hat{H}|\phi \rangle =E|\phi \rangle  \) can easily be solved, giving for
\( n=1,2,\ldots  \), \( |\phi _{n}^{\pm }\rangle =|n\rangle |-\rangle +\sqrt{n}/(\tilde{E}_{n}^{\pm }+\tilde{t})|n-1\rangle |+\rangle  \)
with \( \tilde{E}_{n}^{\pm }=\pm \sqrt{n+\tilde{t}^{2}} \) , and one single
state \( |\phi _{0}\rangle =|0\rangle |-\rangle  \) with \( \tilde{E}_{0}=\tilde{t} \).
Figure \ref{fig:spectre} shows this spectrum with the simplified expressions
of \( |\phi _{n}^{\pm }\rangle  \) obtained for large \( \left| \tilde{t}\right|  \).
These simplified expressions involving the states \( |n\rangle  \) of the harmonic
oscillator express the quantized modes for small oscillations near the extrema
of the two bands \cite{clusterInd}. We clearly observe the exchange of one
state in the spectrum, giving \( \Delta N_{g=\pm }=\pm 1 \). As a consequence,
\( \Delta N_{g}+\Delta C_{g}=0 \). This gives for each band a conserved quantity,
namely \( (N_{g}+C_{g}) \). The variation \( \Delta C_{+}=-1 \) can also be
considered as ``the topological charge'' associated to the degeneracy of the
\( 2\times 2 \) matrix (\ref{e:modele}) where \( a_{cl}=x_{cl}-ip_{cl} \)
is a classical variable. {[}\( \Delta C_{+}=-1 \) is also obtained from Eq.
(\ref{e:formule C+-}), for a little sphere in space \( (x_{cl},p_{cl},\tilde{t}) \)
centered at the degeneracy \( (0,0,0) \){]}.

For a general Hamiltonian \emph{\( \hat{H}({\mathbf{S}},{\mathbf{J}}/j) \)}
the very simple model Eq.(\ref{e:modele}) provides the general mechanism for
the exchange of one state in the vicinity of every degeneracy between two consecutive
bands. In the trivial case \( \hat{H}_{0}=S_{z} \), we have computed the value
of the conserved quantity: \( N_{g}+C_{g}=2j+1 \). We deduce that it is still
correct when \( \hat{H}_{0} \) is deformed to \emph{\( \hat{H}({\mathbf{S}},{\mathbf{J}}/j) \)},
proving formula (\ref{e:formule2}).

The second example, addresses the Chern indices of the two-state model corresponding
to \( s=1/2 \), i.e. to \( \dim {\mathcal{H}}_{s}=2 \). In a given fixed basis
, say \( |m_{s}=\pm 1/2\rangle  \), the matrix of \( \hat{H}_{s}({\mathbf{J}}_{cl}) \)
has the form: 
\begin{equation}
\label{e:2x2matrix}
\hat{H}_{s}\equiv \left( \begin{array}{cc}
h_{11}({\mathbf{J}}_{cl}) & h_{12}({\mathbf{J}}_{cl})\\
\overline{h_{12}}({\mathbf{J}}_{cl}) & h_{22}({\mathbf{J}}_{cl})
\end{array}\right) .
\end{equation}

This matrix will give two vibrational sub-bands with Chern indices \( C_{-} \)
and \( C_{+} \). In Ref.\cite{fred1}, it is shown that the Chern indices have
the following property: Let \( {\mathbf{J}}^{*} \) be the zeros of \( h_{12}({\mathbf{J}}_{cl}) \),
and \( \sigma ({\mathbf{J}}^{*}) \), their degree defined as follows: take
a small direct circle around \( {\mathbf{J}}^{*} \); its image by \( h_{12}({\mathbf{J}}_{cl}) \)
is a closed curve around \( 0 \) with \( \sigma ({\mathbf{J}}^{*}) \) turns.
Define the set \( {\mathcal{S}}^{+}=\left\{ {\mathbf{J}}_{cl}\in S^{2}_{j}\textrm{ such }\, \, \textrm{ that }\left( h_{22}({\mathbf{J}}_{cl})-h_{11}({\mathbf{J}}_{cl})\right) >0\right\}  \).
Then 
\begin{eqnarray}
C_{+}=\sum _{J^{*}\in {\mathcal{S}}^{+}}\sigma ({\mathbf{J}}^{*})=-C_{-}.\label{e:formule C+-} 
\end{eqnarray}

A direct consequence of this property is that a change of Chern index can only
occur when simultaneously \( h_{12}({\mathbf{J}}^{*})=h_{22}({\mathbf{J}}^{*})-h_{11}({\mathbf{J}}^{*})=0. \)
This corresponds to a degeneracy in the spectrum of \( \hat{H}_{s} \), and
a conical contact of the two bands.

Formula (\ref{e:formule C+-}) can be applied to a Hamiltonian describing the
rotational structure of the doubly degenerated vibrational state of a tetrahedral
(or octahedral) spherical top molecule \cite{boris1}. The most general Hamiltonian,
taken up to the third degree in \( \mathbf{J} \), has the form (\ref{e:2x2matrix})
with

\begin{eqnarray*}
h_{12}({\mathbf{J}})=\left( J_{x}^{2}-J_{y}^{2}\right) /j^{2}+iXJ_{x}J_{y}J_{z}/j^{3}, &  & \\
h_{22}({\mathbf{J}})-h_{11}({\mathbf{J}})=\left( 3J_{z}^{2}-j(j+1)\right) /j^{2}, &  & 
\end{eqnarray*}
 with parameter \( X\in \R  \). In this case the set \( {\mathcal{S}}^{+} \)
includes all points around north and south pole for which \( J_{z}>\frac{1}{\sqrt{3}}\sqrt{j(j+1)} \)
or \( J_{z}<\frac{-1}{\sqrt{3}}\sqrt{j(j+1)} \). As \( h_{12}({\mathbf{J}})=0 \)
for simultaneously \( J_{x}=\pm J_{y} \), and \( J_{x}J_{y}J_{z}=0 \) there
are two points \( J^{*}\in {\mathcal{S}}^{+} \): the north pole \( J_{z}=j \)
and the south pole \( J_{z}=-j \). We consider \( {\mathbf{J}} \) going through
a closed path surrounding each \( J^{*} \) to calculate \( \sigma ({\mathbf{J}}^{*}) \).This
gives for north and for the south poles \( \sigma _{north}=\sigma _{south}=2\, \textrm{sign}(X) \),
so that \( C_{+}=4\, \textrm{sign}(X) \) and \( C_{-}=-C_{+} \).

This calculation explains why the rotational structure of doubly degenerate
vibrational state is generally split into two sub-bands with respectively \( 2j+5 \)
and \( 2j-3 \) levels \cite{mich,ZhBr}. In our current approach the appearance
of two bands with Chern indices \( \pm 4 \) for the Hamiltonian (\ref{e:2x2matrix})
is due to the formation of eight degeneracies {[}equivalent by symmetry{]} between
the two vibrational bands at \( X=0 \) parameter value. More generally the
characterization of the rovibrational structure of molecules and its possible
modification under the variation of some physical parameters like total angular
momentum can be done systematically by using Chern indices as topological quantum
numbers. In Ref.\cite{BrZh} similar effect was discussed without explicit introduction
of Chern indices.

The molecular application studied in this Letter was mainly inspired by the
role played by topological Chern indices in the integer quantum Hall effect
\cite{chern3}. In this context, Chern indices describe the topology of Floquet
bands \( (\mathbf{k})\rightarrow [|\psi _{g}(\mathbf{k})\rangle ] \) where
\( \mathbf{k} \) is the Bloch wave vector, and give a quantum Hall conductance
\( \sigma _{g}=(e^{2}/h)C_{g} \) under the hypothesis of adiabatic motion of
\( \mathbf{k} \) when a weak electrical field is applied. Contrary to Eq.(\ref{eq/sum}),
their sum for a given Landau level is \( \sum _{g}C_{g}=+1 \), because of the
non trivial topology of the quantum space \cite{chern3}. Many properties of
the Chern indices are similar: a change \( \Delta C_{g}=\pm 1 \) is related
to a conical degeneracy between consecutive bands. The application of semi-classical
calculations of \( C_{g} \) done for the Hall conductance in Ref.\cite{fred3},
will be the subject of future work.

In summary we have discussed the role of Chern quantum numbers to molecular
spectroscopy. The interpretation of good integer quantum numbers associated
with rotational structure of different vibrational bands in terms of topological
Chern numbers has naturally a wide applicability. These indices can be introduced
any time when the adiabatic separation of variables enables one to split the
global structure into bands associated with the ``fast motion'' and their
internal structure described by a ``slow motion'' on a compact phase space.
We have considered here only the problem when the dimension of the classical
phase space formed by the slow variable \( {\mathbf{J}}_{cl} \) is 2. Only
the first Chern class appears in this case. Extension to higher dimension requires
more delicate physical interpretation and more sophisticated mathematical tools
in relation with the index theorem of Atiyah-Singer. In molecular spectroscopy
many problems with intra-molecular dynamics are known in great detail. They
can be used to test the applicability of this new concept.

We gratefully acknowledge Y. Colin de Verdiere for stimulating discussions and
Bart Van-Tiggelen for helpful comments.

\end{document}